\documentclass[aps,pra,twocolumn,showpacs,longbibliography,superscriptaddress,floatfix]{revtex4-1}
\usepackage[colorlinks=true, allcolors=blue]{hyperref}
\usepackage{amsmath}
\usepackage{dsfont}
\usepackage{amssymb}
\usepackage{braket}
\usepackage{color}
\usepackage{eucal}
\usepackage{xcolor}
\usepackage{subfig}
\usepackage{caption}
\usepackage{graphicx}

\begin{document}
\title{Minimum uncertainty states and squeezed states from sum uncertainty relation}

\author{Yatindra Kumar}
\email{Now at Indian Institute of Science Education and Research, Pune, India}
\affiliation{Department of Physics, Institute of Science, BHU, Varanasi 221005, India}

\author{Yashraj Jha}
\email{Now at Indian Institute of Science Education and Research, Berhampur, India}
\affiliation{Department of Physics, Institute of Science, BHU, Varanasi 221005, India}

\author{Namrata Shukla}
\email{namrata.shukla@bhu.ac.in}
\affiliation{Department of Physics, Institute of Science, BHU, Varanasi 221005, India}

\begin{abstract}
Heisenberg's uncertainty relation is at the origin of understanding minimum uncertainty states and squeezed states of light.~In the recent past, sum uncertainty relation was formulated by Maccone and Pati [Maccone \& Pati, Phys. Rev. Lett. 113, 260401 (2014)] which is claimed to be stronger than the existing Heisenberg-Robertson product uncertainty relation.~We analyze the minimum uncertainty states for the sum uncertainty relation using the variational approach.~We claim that the minimum uncertainty states for the sum uncertainty relation are always the minimum uncertainty states for the traditional product uncertainty relation, using the example of position-momentum pair as well as angular momentum operators.~We show that the coherent and squeezed states of radiation remain completely unaffected by the sum uncertainty relation.
\end{abstract}

\maketitle

\section{Introduction}
Heisenberg's uncertainty relation is the mathematical formulation of one of the most foundational principles of quantum mechanics known as uncertainty principle~\cite{Hei_1927, Rob_1929, Whe_b_1983, Uff_1985, Uff_1988}.~Despite being foundational, uncertainty relations have been debated about their interpretation in the recent decades~\cite{Bus_2007, Hof_2003,Guh_2004, Oza_2003, Roz_2012, Bus_2013, Busc_2014, Bus_2015, Muk_2016, Hub_2018, Kan_2014, Erh_2012, Sul_2017, sul_2013}.~The Robertson's formalized version of the uncertainty relation gives a lower bound on the product of variances for the two arbitrary and incompatible observables.~It was pointed out that the Heisenberg's uncertainty relation can be trivial in some cases as the variances of the observables takes the `null' value.~This gave rise to the formulation of sum uncertainty relations \cite{Mac_2014}.\\
In the field of quantum optics, the traditional uncertainty relation is at the origin of understanding coherent state which is the minimum uncertainty state and saturates the uncertainty relation \cite{Gla_1963}.~Minimum uncertainty states (MUS) are tremendously useful in the field of quantum metrology and sensing \cite{Fra_2021,Mal_2023}.~The MUS led to the squeezed states of light \cite{Wal_1983, Zub_2005, Leo13, Sch01} and the squeezed light not only gets actively used in entanglement detection \cite{Tot_2009,Nat_2002} but also is pivotal to encoding quantum information for secure communication \cite{Big_2001,Tra_2002,Mot_2017,nam_2021,Car_2021,Xu_2019,Shu_2021}.


Although, the concept of MUS is well understood for the traditional product uncertainty relation, it still requires a detailed analysis for the sum uncertainty relation.~In this work, We analyze MUS for the sum uncertainty relation, specially in the quantum optical context.~Although, Maccone and Pati \cite{Mac_2014} talk about the minimization and maximization of the bound for the sum uncertainty relations, they do not discuss MUS in general.~We investigate and discuss the MUS with relevant examples for finite dimensional case as well as continuous variable systems concluding that the coherent and squeezed states remain unchanged due to the sum uncertainty relations.\\

The Heisenberg-Robertson uncertainty relation for two incompatible observables $A$ and $B$ reads  
\begin{equation}
\Delta A^2\Delta B^2\geq \frac{1}{4}{\vert\bra{\psi}[A,B]\ket{\psi}\vert}^2,
\label{Rob-Heis_UR}
\end{equation}    
where $\Delta A^2$ and $\Delta B^2$ are the variances.~Conventionally, the MUS are those satisfying
\begin{equation}
\Delta A\Delta B=\frac{1}{2}{\vert\bra{\psi}[A,B]\ket{\psi}\vert}.
\label{Rob-Heis_MUS}
\end{equation}
It is also well established that MUS \cite{Vin_2021,Kec_2016,Alo_2014,Rom_1968} are eigenstates of the operator $A\pm i\gamma B$ for $\gamma = \Delta A/\Delta B$.\\
For position and momentum operators, the MUS is eigenstate of annihilation operator.~This state is also compared with the ground state wavefunction of the displaced harmonic oscillator more popularly known as coherent state with $\Delta x\Delta p=\frac{\hbar}{2}$.~The quadrature operators $X_1$ and  $X_2$ are defined as 
\begin{equation}
 X_1=\frac{1}{2}\left(a+a^\dagger\right),~X_2=\frac{1}{2 i}\left(a-a^\dagger\right),  \label{Quadrature_operators}
\end{equation} for $[a,a^\dagger]=1$, where $a, a^\dagger$ are annihilation and creation operators, respectively.~These quadrature operators are essentially dimensionless position and momentum operators by definition and follow the uncertainty relation 
\begin{equation}
\Delta X_1 \Delta X_2\geq\frac{1}{4}.
\label{Uncertainty_quadrature}
\end{equation}
The above uncertainty relation gets saturated by the MUS for $\Delta X_1=\Delta X_2=\frac{1}{2}$.~A squeezed state \cite{Zub_2005} is defined as the state for which the variance is smaller than that of the corresponding coherent state, i.e., $
\Delta X_1^2<\frac{1}{4}$
represents squeezed state in $X_1$ quadrature.~The generalization of the concept of squeezed state is also established in the context of polarization squeezing \cite{Ran_2011, Nam_2016} and spin squeezing \cite{Tot_2009} in the field of quantum optics.\\
Maccone and Pati~\cite{Mac_2014} proposed a generalized sum uncertainty relation
\begin{equation}
\Delta A^2+\Delta B^2\geq \pm i\langle[A,B]\rangle+\vert\bra{\psi} A \pm i B \ket{\psi^\perp}\vert^2.
\label{Mac-Pati_UR}
\end{equation}  
The choice of sign is such that the real quantity $\pm i\langle[A,B]\rangle$ takes a positive value.~This inequality is valid for any $\ket{\psi^{\perp}}$ orthogonal to the state $\ket{\psi}$ as long as it is not orthogonal to the state $(A\pm i B)\ket{\psi}$.~Maccone and Pati~\cite{Mac_2014} also proposed another sum uncertainty relation
\begin{equation}
    \Delta A^2+\Delta B^2\geq \frac{1}{2} \vert \bra{\psi^{\perp}_{A+B} }A+B \ket{\psi} \vert^2
    \label{Mac-Pati_UR_2}
\end{equation}
where $\ket{\psi^{\perp}_{A+B}} \propto ((A+B) - \braket{A+B}) \ket{\psi}$.

Maccone and Pati \cite{Mac_2014} clearly state that the maximization of the lower bound for the correct choice of $\ket{\psi^\perp}$ corresponds to the equality.~Notably, maximizing RHS with respect to $\ket{\psi^\perp}$ such that the second term is nonzero is not relevant for finding MUS.~Although, the inequalities by their very nature saturate for the maximum of the RHS, it is not clear what choices of $\ket{\psi}$ saturate the sum uncertainty relation \cite{Mac_2014}.~We attempt to answer this question here.~The paper \cite{Mac_2014} also mentions a simpler lower bound on the sum of variances
\begin{equation}
\Delta A^2+\Delta B^2 \geq \vert \langle [A,B] \rangle\vert.
\label{new_sum_UR}
\end{equation}
We show that this inequality is necessary and sufficient to define the MUS for the sum uncertainty relation\eqref{Mac-Pati_UR} and \eqref{Mac-Pati_UR_2}.\\

With an additional variable $\ket{\psi^\perp}$ in the system, when maximizing the RHS with respect to $\ket{\psi^\perp}$, one can still choose a $\ket{\psi}$ that gives a local minima for the first term in \eqref{Mac-Pati_UR} and make the second term null.~This motivates our approach to use variational method to extremize both the LHS and RHS of \eqref{Mac-Pati_UR} and look for the states $\ket{\psi}$ for which these extrema coincide, saturating the inequality.~We apply this approach to the position-momentum pair and also numerically study the discrete case for the angular momentum operators. \\

\section{Minimum uncertainty states for sum uncertainty relation}
In the quest of MUS for the sum uncertainty relation \eqref{Mac-Pati_UR},~firstly we try to see the role of second term in the RHS involving $\ket{\psi^\perp}$.~We start by defining two vectors $\ket{f}=C\mp i D \ket{\psi}$,~$\ket{g}=\ket{\psi^{\perp}}$, where $C\equiv A-\langle A \rangle$,~$D\equiv B-\langle B \rangle$.~The inequality \eqref{Mac-Pati_UR} can be proved using the Cauchy-Schwarz inequality form \cite{Mac_2014}
\begin{equation}
\langle f \ket{f} \langle g\ket{g}\geq \vert \langle f\ket{g}\vert^2
\label{C-S inequality}
\end{equation}
The equality corresponding to the sum uncertainty relation~\eqref{Mac-Pati_UR} follows from the equality corresponding to the Cauchy-Schwarz inequality\eqref{C-S inequality} if and only if $\ket{f} \propto \ket{g}$ \cite{Mac_2014}.~However, it is straight forward to show that the inner product $\langle\psi|f\rangle=0$ implying that $\ket{f}$ is also a state orthogonal to $\ket{\psi}$.~Given, infinitely many orthogonal states to the state $\ket{\psi}$, if we choose the $\ket{\psi^{\perp}}=\ket{f}=C\mp i D \ket{\psi}$, $\ket{f}\propto \ket{g}$ is guaranteed.~This means, we can always have at least one $\ket{\psi^{\perp}}$ for every state $\ket{\psi}$ which can turn the inequality \eqref{Mac-Pati_UR} into an equality making each state the MUS.\\
It is important to note that the inequality \eqref{new_sum_UR} can also be written as
\begin{equation}
\Delta A^2+\Delta B^2 \geq\pm i \langle [A,B] \rangle, 
\label{new_sum_UR_final}    
\end{equation}
and the RHS of the above equation is the first term of the sum uncertainty relation \eqref{Mac-Pati_UR} for which we move on to analyzing the MUS.
For the observables of interest being position and momentum and $[\hat{x},\hat{p}] = i\hbar$,
the inequality \eqref{new_sum_UR_final} takes the form 
$\Delta x^2 + \Delta p^2 \geq \hbar$,
and the MUS satisfy  
$\Delta x^2 + \Delta p^2 = \hbar$.
It is trivial to show that the above equation is satisfied by the eigenstates of the operator $\hat{x}+i\hat{p}$ with $\Delta p^2=\Delta x^2=\frac{\hbar}{2}$.~The equation is also satisfied for the coherent and squeezed states of radiation. 


We now inspect the extrema of LHS and RHS in \eqref{Mac-Pati_UR} and find the states for which they overlap, to ensure MUS.~In the first step, we use variational technique to extremize the LHS of Eq.~\eqref{Mac-Pati_UR} which can be written in the form
\begin{equation}
LHS= \int \psi^* (A-\braket{A})^2 \psi dx+\int \psi^* (B-\braket{B})^2 \psi dx.
\end{equation}
Taking into account the constraint $\int \psi^*\psi dx =1$, we introduce the Lagrange multiplier $\lambda$ to define
\begin{equation}
J_L=LHS-\lambda\int \psi^* \psi ~dx, 
\label{funct_min_LHS}
\end{equation}
where the variables $\psi$ and $\psi^*$ are treated independently.~To maximize the functional $J_L$, using variational derivatives we enforce the condition $\delta J_L=0$.~On varying \eqref{funct_min_LHS} with respect to $\psi^*$, we get 
\begin{equation}
\frac{\delta J_L}{\delta \psi^*}=(A-\braket{A})^2 \psi + (B-\braket{B})^2 \psi - \lambda \psi = 0.
\label{der_min_LHS}
\end{equation}
On multiplying \eqref{der_min_LHS} by $\psi^*$ and integrating gives
\begin{equation}
\lambda= \int \psi^* (A-\braket{A})^2 \psi dx+\int \psi^* (B-\braket{B})^2 \psi dx    
\end{equation}
 i.e.~$\lambda= \Delta A^2 + \Delta B^2$.~Eq.~\eqref{der_min_LHS} takes the form of eigenvalue equation 
\begin{equation}
    \left( \frac{(A-\braket{A})^2+(B-\braket{B})^2}{\Delta A^2 + \Delta B^2} \right) \psi_L=\psi_L,
    \label{MUS_Eigval_LHS}
\end{equation}
for $\psi=\psi_L$.~This equation must be satisfied to minimize $\Delta A^2 + \Delta B^2$.~In the next step, we extremize the RHS of the sum uncertainty relation \eqref{Mac-Pati_UR}, again by introducing Lagrange multipliers $\lambda_1$ and $\lambda_2$ and varying the quantity
\begin{equation}
J_R=RHS-\lambda_1\int \psi^* \psi dx - \lambda_2\int  (\psi^\perp)^* \psi^\perp dx,  
\end{equation}
where
\begin{align}
    RHS=&\pm i\int \psi^* [A,B] \psi dx \\ \nonumber
         &+ \int \psi^* (A \pm i B) \psi^\perp dx\int (\psi^\perp)^*(A\mp i B)  \psi dx
\end{align}
We choose to vary $J_R$ with respect to $\psi^*$ and $(\psi^{\perp})^*$, in order to maximize it by imposing the conditions 

\begin{equation}
\frac{\delta J_R}{\delta \psi^*}= \frac{\delta J_R}{\delta (\psi^{{\perp}})^*}=0. 
\label{funct_max_RHS}
\end{equation}~We get the equations
\begin{equation}
\begin{split}
\pm i [A,B] \psi&  \\+(A\pm i&B)  \psi^\perp \left( \int (\psi^\perp)^*(A\mp iB) \psi dx \right) - \lambda_1\psi =0, 
\label{der_max_RHS}\end{split}\end{equation}
and
\begin{equation}
(A\mp i B)\psi \left( \int \psi^* (A\pm i B) \psi^\perp dx \right) - \lambda_2\psi^\perp=0.
\label{der_max_RHS 1}
\end{equation}
Multiplying \eqref{der_max_RHS} by $\psi^*$ and \eqref{der_max_RHS 1} by $(\psi^\perp)^*$ and integrating, we get
\begin{align} \nonumber
&\lambda_1=\pm i \langle [A,B] \rangle +\vert\bra{\psi} A \pm iB \ket{\psi^\perp}\vert^2 \\ 
&\lambda_2=\vert\bra{\psi} A \pm iB \ket{\psi^\perp}\vert^2 
\end{align}
Substituting $\lambda_2$ in Eq.~\eqref{der_max_RHS 1} we get
\begin{equation}
\psi^\perp=\frac{(A\mp i B)}{\vert\bra{\psi^\perp} A \mp i B \ket{\psi}\vert} \psi 
\label{MUS_eigval_RHS}
\end{equation}
Furthermore, substituting \eqref{MUS_eigval_RHS} and $\lambda_2$ in Eq.~\eqref{der_max_RHS}, we have the eigenvalue equation
\begin{equation}
    \left( \frac{A^2 + B^2}{ \pm i \langle [A,B] \rangle+\vert\bra{\psi} A \pm iB \ket{\psi^\perp}\vert^2} \right) \psi_R=\psi_R,
\label{MUS_eigval_RHS_1}
\end{equation}
gor $\psi=\psi_R$.~Although, the solutions of this equation may technically represent maxima, minima or inflection points of
$J_L$, for our purposes as mentioned in the previous section, the 
MUS of the sum uncertainty relation \eqref{Mac-Pati_UR} must satisfy Eq.~\eqref{MUS_Eigval_LHS} and Eq.~\eqref{MUS_eigval_RHS_1}.~The wavefunctions $\psi_L$ and $\psi_R$ coincide for $\langle A\rangle=\langle B\rangle=0$, corresponding to the MUS for the sum uncertainty relation \eqref{Mac-Pati_UR}.\\

\section{Position-momentum and harmonic oscillator}
Let us choose the observables  $A=\hat{x}$ and $B=\hat{p}=-i\hbar\frac{d}{dx}$ in the position basis.~In order to extremize the LHS for the MUS, the eigenvalue equation \eqref{MUS_Eigval_LHS} takes the form
\begin{equation}
    \left( \frac{(x-a)^2+(-i\hbar\frac{d}{dx}-b)^2}{m} \right) \psi_L(x)=\psi_L(x),
\end{equation}
where, $a=\braket{x}$, $b=\braket{p}$, $m = \Delta x^2 + \Delta p^2 $.~This leads to the differential equation 
\begin{equation}
\frac{d^2\psi_L}{dx^2}- \frac{2i b}{\hbar}\frac{d\psi_L}{dx}- \frac{m}{\hbar^2}\left(\frac{b^2}{m}+\frac{(x-a)^2}{m}-1\right)\psi_L=0.
\label{diff_eq_LHS_XP}
\end{equation}
We now introduce $u_1(x)=\frac{d\ln{\psi_L}}{dx}$ which gives 
\begin{equation}
\psi_L(x)= \mathcal{N}_L\exp{\int^{x} u_1(\mu)d\mu},  
\label{sol_psi(x)_LHS}
\end{equation}where $\mathcal{N}_L$ is an arbitrary constant of integration.~One gets the following differential equation in $u_1(x)$
\begin{equation}
    \frac{du_1}{dx}+u_1^2-\frac{2i b}{\hbar}u_1= \frac{m}{\hbar^2}\left(\frac{b^2}{m}+\frac{(x-a)^2}{m}-1\right).
    \label{Riccati_LHS}
\end{equation}
This equation is the first order Riccati differential equation.~One can guess a solution of the form $u_1=C_1x+C_2$, where $C_1$ and $C_2$ are complex constants.~By putting in this solution in Eq.~\eqref{Riccati_LHS} and comparing the coefficients, we get 
$C_1=-\frac{1}{\hbar},  C_2=\frac{a}{\hbar}+\frac{i b}{\hbar}$ and $ m=\hbar$ and the solution is
\begin{equation}
    u_1(x)=-\frac{x}{\hbar}+\frac{a}{\hbar}+\frac{i b}{\hbar}.
    \label{Sol_u(x)_LHS}
\end{equation}
On using this in Eq.~\eqref{sol_psi(x)_LHS} 
\begin{equation}
    \psi_L(x)=N_L \exp \left[-\frac{1}{2\hbar}x^2+\left( \frac{a+i b}{\hbar}\right)x \right],
    \label{Final_Gauss_LHS}
\end{equation}
The above wave function $\psi_L(x)$ is the equation to be satisfied by the state that minimizes the LHS of the sum uncertainty relation \eqref{Mac-Pati_UR}.~Applying the same approach to extremize the RHS for MUS, Eq.~\eqref{MUS_eigval_RHS_1} reads
\begin{equation}
\left( \frac{x^2+(-i\hbar\frac{d}{dx})^2}{m^\prime} \right)\psi_R(x)=\psi_R(x),
\end{equation}    
where, $m^\prime = \hbar+\vert\bra{\psi} \hat{x} - i\hat{p} \ket{\psi^\perp}\vert^2$.~It can also be written as
\begin{equation}
(x^2-m^\prime)\psi_R(x)=\hbar^2\frac{d^2\psi_R}{dx^2}.
\end{equation}
We perform the same procedure as in minimizing the LHS by introducing $u_2(x)=\frac{d\ln{\psi_R}}{dx}$ implying 
\begin{equation}
\psi_R(x)=\mathcal{N}_R \exp{\int^{x} u_2(\nu)d\nu}, 
\label{sol_psi(x)_RHS}
\end{equation}
$\mathcal{N}_R$ being an arbitrary constant of integration.~The differential equation in $u_2(x)$ reads
\begin{equation}
\frac{du_2}{dx}+u_2^2=\frac{x^2-m^\prime}{\hbar^2},
\label{Riccati_RHS}
\end{equation}
which is the first order Riccati differential equation.~We again guess a feasible solution to be $u_2=C_3x+C_4$ and substitute in the above equation.~We find $C_3=-\frac{1}{\hbar},~C_4=0$ and $m^\prime=\hbar$.~This leads to 
\begin{equation}
u_2(x)=-\frac{x}{\hbar},  
\label{sol_u(x)_RHS}
\end{equation}
and substituting this in \eqref{sol_psi(x)_RHS} gives
\begin{equation}
\psi_R(x)=\mathcal{N}_R \exp \left[-\frac{x^2}{2\hbar} \right]
\label{Final_Gauss_RHS}
\end{equation}
The wave functions in Eq.~\eqref{Final_Gauss_LHS} and \eqref{Final_Gauss_RHS} are both minimum uncertainty Gaussians as expected.~However, the two equations are satisfied by the same state only for $\langle x \rangle=\langle p \rangle=0$ which is by definition true for the harmonic oscillator wave functions.~It is pivotal to notice here that $m=m^\prime=\hbar$, meaning the second term of the sum uncertainty relation \eqref{Mac-Pati_UR} $\vert\bra{\psi} A \pm i B \ket{\psi^\perp}\vert^2$ does not contribute when it comes to the MUS for the case of position-momentum.\\
It is worth noting that the RHS for the inequality \eqref{Mac-Pati_UR_2} can be written in the form
\begin{equation}
\begin{split}
\frac{1}{2} \bra{\psi} (A+B)^2 - \braket{A+B}^2 \ket{\psi} = \frac{1}{2} \Delta (A+B)^2
    \end{split}
\end{equation}
with the restricted choice of $\ket{\psi_{A+B}^\perp}$.~Doing the same analysis by extremizing RHS of \eqref{Mac-Pati_UR_2} for the position and momentum leads to the wave function \begin{equation}
    \psi_{\pm} = C_{\pm} \exp{\left( \frac{i}{\hbar} \left(-\frac{x^2}{2} + (\mu \pm \delta)x \right) \right)}
\end{equation}
where $\mu = \braket{A+B}$, $\delta = \Delta(A+B)$.~This wave function is non-normalizable meaning no physical MUS exists for \eqref{Mac-Pati_UR_2}.

\section{Commentary on MUS}
\begin{figure}[htbp]
    \centering
    \subfloat[]{%
        \includegraphics[width=0.45\linewidth]{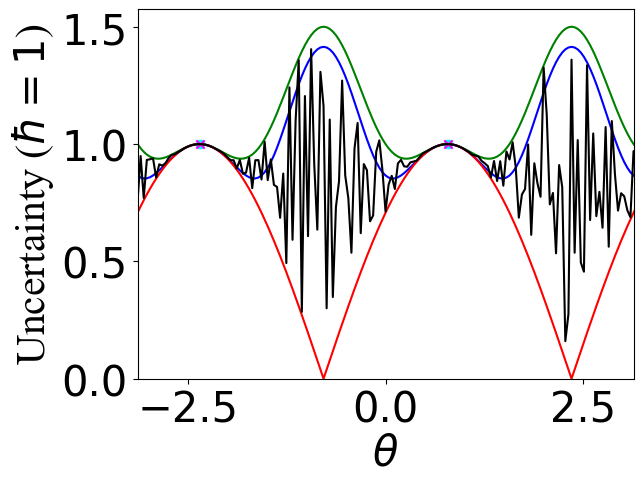}%
        \label{fig:1_perp_psi}
    }
    \hfill
    \subfloat[]{%
        \includegraphics[width=0.45\linewidth]{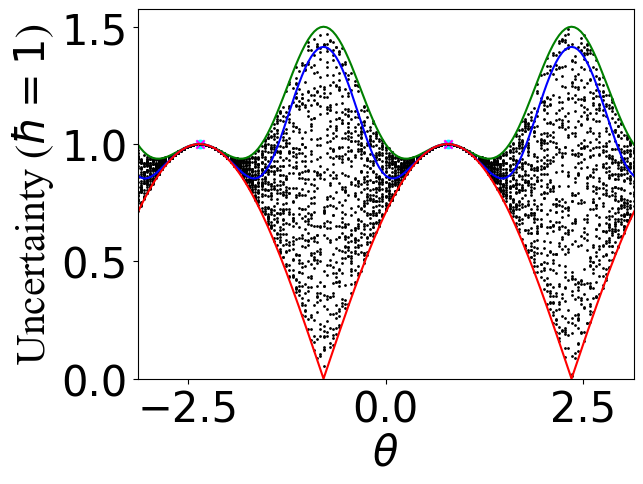}%
        \label{fig:30_perp_psi}
    }
    \caption{$\theta$ dependence for the observables $J_z$ and $J_y$ for the state $\ket{\psi}=\frac{1}{\sqrt{2}}\left[\cos{\theta}\ket{1}+\sin{\theta}\ket{-1}+\ket{0}\right]$, where $\ket{1}$, $\ket{-1}$ and $\ket{0}$ are the eigenstates of $J_z$ operator.~The red curve is the first term of the RHS of inequality \eqref{Mac-Pati_UR}, the green is the LHS of inequality \eqref{Mac-Pati_UR}, the blue curve is the value $2\Delta A \Delta B$, the cyan dots are the eigenstates of the operator $A \mp i B$, and magenta crosses are the eigenstates of operator $A^2+B^2$.~RHS of the inequality \eqref{Mac-Pati_UR} is represented by (a) black lines for one random $\ket{\psi^{\perp}}$ for each $\theta$ value, and (b) black dots for thirty random $\ket{\psi^{\perp}}$ for each $\theta$ value.~The $\ket{\psi^\perp}$ here is generated using Haar random unitary.}
    \label{fig:Jz_Jy_plot}
\end{figure}

As shown in the last subsection by variational method for the position and momentum, the second term of the sum uncertainty relation \eqref{Mac-Pati_UR} may not be relevant when it comes to the MUS.~Indeed, the states saturating the inequality \eqref{new_sum_UR} will be the MUS for \eqref{Mac-Pati_UR}.~It is obvious at this point that these MUS will also be MUS for the traditional product uncertainty relation \eqref{Rob-Heis_UR} following the minimization of the LHS between \eqref{Rob-Heis_UR} and \eqref{new_sum_UR} because the RHS is constant.\\
We further test and verify this result for the angular momentum operators $J_y$ and $J_z$ (see Fig.~\ref{fig:Jz_Jy_plot}) by randomly generating $\ket{\psi^\perp}$ in the carefully parametrized state space to capture the MUS.~The random $\ket{\psi^\perp}$ here is generated by using a Haar uniform randomly generated unitary $U$ \cite{Kar_1998}, and $\ket{\psi^\perp}\propto\left(\mathds{1}-\ket{\psi}\bra{\psi}\right)U\ket{1}$.~We choose the observables $J_z$ and $J_y$ and randomly generate one or many $\ket{\psi^\perp}$ for the states $\ket{\psi} =\frac{1}{\sqrt{2}}\left[\left(\cos{\theta}\ket{1}+\sin{\theta}\ket{-1}\right)+\ket{0}\right]$ to demonstrate the following points.~First, the second term $\vert\bra{\psi} A \pm i B \ket{\psi^\perp}\vert^2$ in the sum uncertainty relation \eqref{Mac-Pati_UR} does not play into the MUS.~Second, the MUS for \eqref{Mac-Pati_UR}, are certainly the MUS for \eqref{new_sum_UR} and also \eqref{Rob-Heis_UR}.~Third, the MUS are eigenstates of not only $A\pm iB$ but also the operator $A^2+B^2$.~We also execute this numerical verification in higher dimensions e.g., for a superposition of four spin eigenstates spanning the Hilbert space.~As expected, our results hold good independent of the Hilbert space structure as long as the algebra is intact.\\
~The two cases studied here suggest that, even though \eqref{new_sum_UR} is a weaker inequality, it suffices to saturate it for finding the MUS implying $\vert\bra{\psi} A \pm i B \ket{\psi^\perp}\vert^2=0$.~Therefore, the MUS are the eigenstates of the operator $A\pm i B$ which is a special case $(\gamma= 1)$ of $A\pm i \gamma B$.
\\
It is obvious that, the sum uncertainty relation \eqref{new_sum_UR} for the quadrature operators $X_1$ and $X_2$ takes the form $\Delta X_1^2+\Delta X_2^2\geq\frac{1}{2}$.~The corresponding equality satisfied by the MUS with $\Delta X_1^2=\Delta X_2^2=\frac{1}{4}$ and the squeezed state of radiation is obtained for $\Delta X_1^2<\frac{1}{4}$.~This is no different than the the existing definition of squeezed states.\newline
As far as the inequality \eqref{Mac-Pati_UR_2} is concerned, there is no intersections of LHS and RHS seen for the above settings which is completely in agreement with the analytically proved case of of position and momentum.
\section{Conclusion}
We analyze the MUS for the sum uncertainty relation \eqref{Mac-Pati_UR} and find that the inequality \eqref{new_sum_UR} mentioned by Maccone and Pati is necessary and sufficient to define the MUS.~We use variational method for the position-momentum pair to find the MUS for the sum uncertainty relation \eqref{Mac-Pati_UR}.~The example suggests that such a state is coherent state, i.e., the ground state of a harmonic oscillator.~Additionally, we numerically show for the angular momentum operators that the MUS of sum uncertainty relation \eqref{Mac-Pati_UR} are also the MUS of the traditional product uncertainty relation \eqref{Rob-Heis_UR}.~Our main result is that if one were to define the coherent state and squeezed states of radiation from the sum uncertainty relation \cite{Mac_2014}, it would still be the same as defined for the product uncertainty relation \eqref{Rob-Heis_UR}.\\
\paragraph*{Acknowledgements}
NS acknowledges the IOE seed grant by the Banaras Hindu University
(Seed Grant II/2021-22/39995).~The authors acknowledge Debanand Sa, V S Subrahmanyam and Pratap Singh for fruitful discussions.

\bibliography{references}
\end{document}